\documentclass[graybox]{svmult}

\usepackage{type1cm}        % activate if the above 3 fonts are
\usepackage{makeidx}         % allows index generation
\usepackage{graphicx}        % standard LaTeX graphics tool
\usepackage{multicol}        % used for the two-column index
\usepackage[bottom]{footmisc}% places footnotes at page bottom

\usepackage{newtxtext}       % 
\usepackage{newtxmath}       % selects Times Roman as basic font

\usepackage{booktabs}
\usepackage{multirow}

\usepackage{xcolor} % colored tables

\usepackage{graphicx}% Include figure files
\usepackage{dcolumn}% Align table columns on decimal point
\usepackage{bm}% bold math
\usepackage[flushleft]{threeparttable} % table footnote

\usepackage{dsfont}
\usepackage{color,psfrag}
\usepackage{mathtools,amscd}

\usepackage{soul}
\usepackage{xcolor} % colored tables

\usepackage{color}
\definecolor{lightgray}{gray}{0.75}
\newcommand\greybox[1]{%
  \vskip\baselineskip%
  \par\noindent\colorbox{lightgray}{%
    \begin{minipage}{\textwidth}#1\end{minipage}%
  }%
  \vskip\baselineskip%
}

\usepackage{hyperref}

\hypersetup{
	colorlinks,
	linkcolor=blue,
	citecolor=blue, 
	filecolor=black,
	urlcolor=blue}

\newcommand{\mat}[1]{\mathbf{#1}} % matrices in bold

\newcommand{\tran}{^{\mathstrut\scriptscriptstyle\top}} 
\makeindex             % used for the subject index
\begin{document}

\title*{Accurate Molecular Dynamics Enabled by Efficient Physically-Constrained Machine Learning Approaches}
\author{Stefan Chmiela$^{a}$, Huziel E. Sauceda$^{b}$, Alexandre Tkatchenko$^{c}$ and Klaus-Robert M\"uller$^{a,d,e}$}
\institute{Stefan Chmiela, \email{stefan@chmiela.com}
\at $^a$Machine Learning Group, Technische Universit\"at Berlin, 10587 Berlin, Germany\\
$^b$Fritz-Haber-Institut der Max-Planck-Gesellschaft, 14195 Berlin, Germany\\
$^c$Physics and Materials Science Research Unit, University of Luxembourg, L-1511 Luxembourg\\
$^d$Department of Brain and Cognitive Engineering, Korea University, Anam-dong, Seongbuk-gu, Seoul 02841, Republic of Korea\\
$^e$Max Planck Institute for Informatics, Stuhlsatzenhausweg, 66123 Saarbr\"ucken, Germany}
%

\maketitle

\abstract{We develop a combined machine learning (ML) and quantum mechanics approach that enables data-efficient reconstruction of flexible molecular force fields from high-level \emph{ab initio} calculations, through the consideration of fundamental physical constraints. We discuss how such constraints are recovered and incorporated into ML models. 
Specifically, we use conservation of energy -- a fundamental property of closed classical and quantum mechanical systems -- to derive an efficient gradient-domain machine learning (GDML) model. The challenge of constructing conservative force fields is accomplished by learning in a Hilbert space of vector-valued functions that obey the law of energy conservation.
We proceed with the development of a multi-partite matching algorithm that enables a fully automated recovery of physically relevant point-group and fluxional symmetries from the training dataset into a symmetric variant of our model. 
The developed symmetric GDML (sGDML) approach is able to faithfully reproduce global force fields at the accuracy high-level \emph{ab initio} methods, thus enabling sample intensive tasks like molecular dynamics simulations at that level of accuracy.}

%NOTE: We show how to go from data abundance and no weak priors to data scarcity and strong priors.

\section{Introduction}

Molecular dynamics (MD) simulations have become an indispensable atomistic modeling tool, revealing the equilibrium thermodynamic and dynamical properties of a system, while simultaneously providing atomic-scale insight. The predictive power of such simulations is crucially determined by the accuracy of the underlying description of inter-atomic forces, which is typically the limiting factor. Most commonly, the inter-atomic forces are obtained from classical potentials, which provide a mechanistic description in terms of fixed interaction patterns between bonds and bond angles within a molecule. In contrast to exact \emph{ab initio} methods, classical force fields are computationally affordable enough to allow for long simulation time scales, but their simplistic characterisation of atomic interactions prevents them from capturing wide range of important effects, such as the anharmonic nature of atomic bonds, charge transfer and many-body effects~\cite{tuckerman2002ab}.

As an alternative, a series of methodological advances in the field of machine learning (ML)~\cite{rupp2012fast,Hansen2013,Hansen2015,Rupp2015,botu2015adaptive,Hirn2015,deltaML2015,Ceriotti2016,artrith2017efficient,Ceriotti2017,Glielmo2017,yao2017many,john2017many,faber2017prediction,eickenberg2018solid,glielmo2018efficient,tang2018atomistic,Grisafi2018,pronobis2018many,Behler2007,Bartok2010,Behler2012,Bartok2013,Montavon2013a,Bartok2015,Ramprasad2015,Tristan2015,DeVita2015,Behler2016,Brockherde2017,Gastegger2017,Schutt2017,schutt2017schnet,SchNet2018,huang1707dna,huan2017universal,Shapeev2017,dral2017structure,zhang2018deep,lubbers2018hierarchical,ryczko2018convolutional,kanamori2018exploring,hy2018predicting,Clementi2018,Tristan,noe2018,noe2018b,thomas2018tensor,smith2018less,gubaev2018machine,FCHL2018,christensen2019operators,winter2019,gubaev2019accelerating} have yielded universal approximators with virtually no inherent flexibility restrictions. Parametrized from high-level \emph{ab initio} reference calculations, they are theoretically able to represent arbitrarily accurate force fields that faithfully capture key quantum effects, as represented in the training data. However, the major handicap of such off-the-shelf solutions is that they require extraordinary amounts of computationally expensive reference calculations until convergence to a useful level of predictive robustness. This limits their practicality, as the full procedure of data generation, training and inference has to outpace the reference method that the ML model is based on. Only then do ML models represent a meaningful addition to existing approaches.

This predicament suggests that a tight integration between ML and fundamental concepts from physics is necessary to close the gap between efficient FFs and accurate high-level ab initio methods. To this end, we review ideas of how to take advantage of conserved quantities in dynamical processes in addition to other physical laws in order to inform universal approximators without compromising their generality. In doing so, we focus statistical inference on the challenging aspects of the problem, while readily available \emph{a priori} knowledge about the atomic interactions is represented exactly and artifact-free through explicit constraints. A key philosophical difference distinguishes this approach from classical FFs: we aim to exclude physically impossible interactions, as opposed to (approximately) parametrizing known behaviour. Inherent modelling biases can thus be avoided. It is a fascinating prospect, that the necessary constraints can often be expressed in simple terms, although they originate from complex interactions, as many regularities can be exploited without an explicit concept of the underlying principles that cause them. For example, Noether's theorem~\cite{noether1918invarianten} allows us to derive symmetry constraints as conservation laws from the Lagrangian.

We approach this challenge using principles of probabilistic inference, which define a set of hypotheses and conditions them on the made observations. The resulting predictions are particularly robust to over-fitting, because all viable hypotheses that agree with the data are taken into account. The incorporation of strong priors restricts this set of hypotheses, which carries data efficiency advantages over approaches that start from a more general set of assumptions. So-called Hilbert space learning algorithms~\cite{Muller2001} will enable us to rigorously incorporate fundamental temporal and spatial symmetries of atomic systems to create models for which parametrization from highly accurate, but costly coupled cluster reference data becomes viable. In the end, we will be able to perform sampling intensive, long-time scale path-integral MD at that level of accuracy.

\section{Hilbert space learning}

Supervised ML infers a relationship between pairs of inputs $\vec{x} \in \mathcal{X}$ and associated outputs $y \in \mathcal{Y}$ from a finite \emph{training set} of $M$ examples. The objective is to formulate a hypothesis that generalizes beyond these known data points, which is estimated by measuring the prediction error of the model on an independent \emph{test set}. While data efficiency, in the sense of a quickly converging generalization error with growing training set size, is certainly a consideration, it has rarely been the primary focus in typical ML application domains. These fields are typically fortuned with an abundance of data and the need for ML purely arises from the lack of theory to describe the sought-after mapping. For example, there is no rigorous way of deriving the contents of an image from its pixel values and the best results are currently achieved by deep neural networks that match each input against hundreds of thousands of characteristic patterns for the respective image class. Here, it does not matter that several millions of example images were needed until the model was able to give useful predictions, because it is the best tool available.

However, the data efficiency demands are much more stringent in the sciences where the baseline is an essentially exact theory and the role of ML algorithms is to forego some of the computational complexity by means of empirical inference. Naturally, this also means that a considerable computational cost is associated with the generation of high-level \emph{ab initio} training data in that case, making it rare. Practically obtainable datasets of that quality are thus often too small to enable deep learning architectures to play up their strengths.

A more efficient alternative is provided by Hilbert space learning algorithms, as they operate in spaces of functions that match prior beliefs about the observed process. While a small number of training points is not enough to condition general estimators adequately, it might be sufficient to constrain a well-behaved, physically meaningful function space. This is alluring, because even complex physical processes involve quantities with well understood properties that can be exploited to define the structure of those Hilbert spaces. In the following, we briefly review the fundamentals of Hilbert space learning and highlight different ways of incorporating prior knowledge.

\subsection{Hilbert spaces}

A \emph{Hilbert space} $\mathcal{H}$ is a vector space over $\mathbb{R}$ with an inner product that yields a complete metric space. The inner product gives rise to a norm $\| \vec{x} \| = \sqrt{\langle \vec{x},\vec{x} \rangle}$, which induces a distance metric $d(\vec{x},\vec{x}') = \| \vec{x} - \vec{x}' \|$, for $\vec{x},\vec{x}'  \in \mathcal{H}$.  Although any $N$-dimensional Euclidean space $\mathbb{R}^N$ is technically a Hilbert space, this formalism becomes particularly interesting in infinite dimension, where $\mathcal{H}$ is a space of functions, while retaining almost all of linear algebra from vector spaces~\cite{scholkopf2001learning, Muller2001}.

\subsubsection{Reproducing kernels}

Many ML algorithms make use of infinite dimensional Hilbert spaces indirectly via the so-called \emph{kernel-trick}, which allows to express inner products of mappings $\Phi: \mathcal{X} \rightarrow  \mathcal{H}$ in terms of inputs $\vec{x} \in \mathcal{X}$ via a \emph{kernel function} $k:\mathcal{X} \times \mathcal{X} \rightarrow \mathbb{R}$:
\begin{align}
k(\vec{x}, \vec{x}') = \langle \Phi(\vec{x}), \Phi(\vec{x}') \rangle_\mathcal{H} \text{.}
\label{eq:kernel_definition}
\end{align}
Eq.~\ref{eq:kernel_definition} holds true for any symmetric and positive semi-definite kernel, i.e. it is required that $k(\vec{x}, \vec{x}') = k(\vec{x}', \vec{x})$ and any linear combination $f = \sum_i \alpha_i \Phi(\vec{x}_i)$ with $\alpha_i \in \mathbb{R}$ must satisfy
\begin{align}
\langle f, f \rangle_\mathcal{H} = \sum_{ij} \alpha_i \alpha_j k(\vec{x}_i, \vec{x}_j) \geq 0 \text{.}
\end{align}
These two properties guarantee the \emph{reproducing property} of $\mathcal{H}$
\begin{align}
f(\vec{x}) = \langle k(\cdot,\vec{x}), f \rangle_\mathcal{H} \text{,}
\end{align}
due to which any evaluation of $f$ corresponds to an inner product evaluation in $\mathcal{H}$ between the representer $k(\cdot,\vec{x}) = \Phi(\vec{x})$ of $\vec{x}$ and the function itself. We say that $k$ is reproducing for a subset of $\mathcal{H}$, the \emph{reproducing kernel Hilbert space} (RKHS). Intuitively, this means that the feature maps $\Phi(\vec{x}_i)$ for all training points $i \in [1, \dots, M]$ provide an over-complete basis for the RKHS~\cite{scholkopf2001learning}.

\subsubsection{Representer theorem}

The computational tractability of Hilbert space learning algorithms is afforded by the \emph{representer theorem} which states that in an RKHS $\mathcal{H}$, the minimizer $\hat{f} \in \mathcal{H}$ of a loss function $\mathcal{L}: \mathcal{Y} \times \mathcal{Y} \rightarrow \mathbb{R}$ in a regularized risk functional with $\lambda > 0$,
% R_\text{reg}[f] = 
\begin{align}
\hat{f} = \underset{f \in \mathcal{F}}{\operatorname{\text{arg min}}}    \left[ \frac{1}{M} \sum^M_i \mathcal{L}(f(\vec{x}_i), y_i) + \lambda \|f\|^2 \right],
\label{eq:regularized_risk_functional}
\end{align}
 admits a representation of the form
\begin{align}
f(\cdot) = \sum^M_i \alpha_i k(\cdot, \vec{x}_i)
\end{align}
for any $\alpha_i$. It therefore reduces the infinite-dimensional minimization problem in a function space to finding the optimal values for a $M$-dimensional vector of coefficients $\vec{\alpha}$~\cite{wahba1990spline, scholkopf2001generalized,Muller2001}. Because we are not fitting a model with a fixed number of predetermined parameters, Hilbert space algorithms are generally regarded as non-parametric methods, i.e. the complexity of the model is able to grow with the amount of available data.

\subsection{Gaussian process models}

When formulated in terms of the squared loss $\mathcal{L}(\hat{f}(\vec{x}), y) = (\hat{f}(\vec{x}) - y)^2$, the regularized risk functional in Eq.~\ref{eq:regularized_risk_functional} can be interpreted as the maximum \emph{a posteriori} estimate of a Gaussian process (GP)~\cite{Muller2001}. One common perspective on GPs is that they specify a prior distribution over a function space. GPs are defined as a collection of random variables that jointly represent the distribution of the function $f(\vec{x})$ at each location $\vec{x}$ and thus as a generalization of the Gaussian probability distribution from vectors to functions. This conceptual extension makes it possible to model complex beliefs.

At least in part, the success of GPs -- in contrast other stochastic processes -- can be attributed to the fact that they are completely defined by only the first- and second-order moments, the mean $\mu(\vec{x})$ and covariance $k(\vec{x},\vec{x}')$ for all pairs of random variables~\cite{rasmussen2004gaussian}:
\begin{align}
f(\vec{x}) \sim \mathcal{GP} \left[\mu(\vec{x}), k(\vec{x},\vec{x}') \right] \text{.}
\end{align}
Any symmetric and positive definite function is a valid covariance that specifies the prior distribution over functions we expect to observe and want to capture by a GP. Altering this function can change the realizations of the GP drastically: e.g. the squared exponential kernel $k(\vec{x},\vec{x}') = \exp(-\|\vec{x} -\vec{x}'\|^2 \sigma^{-1})$ (with a freely selectable length-scale parameters $\sigma$) defines a smooth, infinitely differentiable function space, whereas the exponential kernel $k(\vec{x},\vec{x}') = \exp(-\|\vec{x} -\vec{x}'\| \sigma^{-1})$ produces a non-differentiable realizations. The ability to define a prior explicitly, gives us the opportunity to express a wide range of hypotheses like boundary conditions, coupling between variables or different symmetries like periodicity or group invariants. Most critically, the prior characterizes the generalization behavior of the GP, defining how it extrapolates to previously unseen data. Furthermore, the closure properties of covariance functions allow many compositions, providing additional flexibility to encode complex domain knowledge from existing simple priors~\cite{duvenaud2014automatic}.

The challenge in applying GP models lies in finding a kernel that represents the structure in the data that is being modeled. Many kernels are able to approximate universal continuous functions on a compact subset arbitrarily well, but a strong prior restricts the hypothesis space and drastically improves the convergence of a GP while preventing overfitting~\cite{micchelli2006universal}. Each training point conditions the GP, which allows increasingly accurate predictions from the posterior distribution over functions with growing training set size.

A number of attractive properties beyond their expressivity make GPs particularly useful in the physical sciences: 
 \begin{itemize}
	\item There is a unique and exact closed form solution for the predictive posterior, which allows GPs to be trained analytically. Not only is this faster and more accurate than numerical solvers, but also more robust. E.g. choosing the hyper-parameters of the numerical solver for NNs often involves intuition and time-consuming trial and error.
	\item Because a trained model is the average of \emph{all} hypotheses that agree with the data, GPs are less prone to overfit, which minimizes the chance of artifacts in the reconstruction~\cite{damianou2013deep}. Other types of methods that start from a more general hypothesis space require more complex regularization schemes.
	\item Lastly, their simple linear form makes GPs easier to interpret, which simplifies analysis of the modeled phenomena and supports theory building.
 \end{itemize}

\subsubsection{Gaussian process regression}

It is straightforward to use GPs for regression: Given a sample $(\mat{X}, \vec{y}) = \{(\vec{x}_i, y_i)\}^M_{i}$, we compute the sample covariance matrix $(\mat{K})_{ij} = k(\vec{x}_i, \vec{x}_j)$ and use the posterior mean
\begin{align}
\mu(\vec{x}) = \mathbb{E}[f(\vec{x})] = k_{\mat{X}}(\vec{x})\tran (\mat{K} + \lambda \mathbb{I})^{-1} \vec{y}
\end{align}
to make predictions about new points $\vec{x}$. Here, $k_{\mat{X}}(\vec{x}) = [k(\vec{x},\vec{x}_1), \dots, k(\vec{x},\vec{x}_M)]\tran$ is the vector of covariances between the new point $\vec{x}$ and all training points. In the frequentist interpretation, this algorithm is also referred to as \emph{Kernel Ridge Regression}.

We can also calculate the variability of the hypotheses at every point via the posterior variance
\begin{align}
\sigma^2(\vec{x}) = \mathbb{E}\left[\left(f(\vec{x}) - \mu(\vec{x})\right)^2\right] = k(\vec{x}, \vec{x}) - k_{\mat{X}}(\vec{x})\tran(\mat{K} + \lambda \mathbb{I})^{-1}k_{\mat{X}}(\vec{x}) \text{,}
\end{align}
which gives us an idea about the uncertainty of the prediction. We remark here, that the posterior variance is generally not a measure for the accuracy of the prediction. It rather describes how well the hypothesized space of solutions is conditioned by the observations and whether the made assumptions are correct.

\section{Encoding prior knowledge}

Prior knowledge about the problem at hand is an essential ingredient to the learning task, as it can drastically increase the efficiency of the training process and robustness of the reconstruction. A ML model that starts from weak assumptions will require more training data to achieve the same performance, compared to one that is restricted to solutions with certain known properties. A unique feature of GPs is that they provide a direct way to incorporate such constraints on the hypothesis space~\cite{scholkopf2001learning}.

In the context of this chapter, we are particularly interested in regularities that arise from invariances and symmetries of physical systems. Sure enough, the idea to reduce equations in a way that leaves them invariant is not new in physics. In fact, Jacobi already developed a procedure to simplify Hamilton's dynamical equations of mechanics based on the conserved quantities of dynamical systems~\cite{Lanczos1949} in the middle of the eighteenth century. Heisenberg was the first to apply group theory to quantum mechanics, were he exploited the permutational symmetry of indistinguishable quantum particles in 1926. Even in modern physics, new symmetries are still routinely discovered~\cite{brading2003symmetries}.

Here, we will review the three most important ways to include prior knowledge into GPs: via the representation of the input, the construction of suitable mean and the covariance functions. As the choice of covariance function is especially important, which is why we will describe several distinct ways to construct them.

\subsection{Representation}

Once the data is captured, it needs to be represented in terms of features that are considered to be particularly informative, i.e. well-correlated with the predicted quantity. For example, parametrizing a molecular graph in terms of dihedral angles instead of pairwise distances might be advantageous when modeling complex transition paths.

The representation of the data also provides the first opportunity to incorporate known invariances of the task at hand. 
Especially in physical systems, there are transformations that leave its properties invariant, which introduces redundancies that can be exploited with a representation that shares those symmetries.

%HES:
%In particular, the symmetries of the system’s Hamiltonian provide straightforward analytical expressions to be algebraically imposed in the ML model to reformulate the learning problem into a simpler, but equivalent one.
%Transformations that leave the Hamiltonian invariant are, for example, rotations, translations, reflections or atomic permutations. There is another fundamental symmetry that must be fulfilled by the systems of interest in this study, the time invariance of the Hamiltonian. This invariance imposes energy conservation in the system~\cite{BookChap_HES}. 

E.g. physical systems can generally be translated and rotated in space without affecting their attributes. Often, the invariances extend to more interesting group of transformations like rotations, reflections or permutations, providing further opportunities to reformulate the learning problem into a simpler, but equivalent one. Conveniently, any non-linear map $\mat{D}: \mathcal{X} \rightarrow \mathcal{D}$ of the input to a covariance function yields another valid covariance function, providing a direct way to incorporate desired invariances into existing kernels~\cite{mackay1998introduction}.

\subsection{Covariance function}

Symmetries in the input data naturally translate to symmetries in the output. If a molecular graph is mirror symmetric, so will be its potential energy surface. However, sometimes there is structure in the output that is not tied to the input at all. This is the case when the predicted property is subject to a conservation law, e.g. the energy of a system is conserved as its geometry propagates through time. There is no representation of individual data points that would be able to capture this kind of symmetry. 

Instead, conservation laws have to be incorporated as constraints into the predictor, to restrict the space of eligible solutions. This is achieved elegantly in GPs, via modification of the covariance function in a way that gives rise to a prior that obeys the desired symmetry. Any function drawn from that prior will then inherit the same invariances~\cite{scholkopf2001learning, rasmussen2004gaussian}. Before developing a covariance function that fits our problem, we will briefly highlight different ways to construct them. After all, arbitrary functions of two inputs $\vec{x},\vec{x}' \in \mathcal{X}$ are not necessarily valid covariance functions. For that purpose we will switch away from the probabilistic view that we held so far and provide a perspective that is more intuitive in the physics context. 

\subsubsection{Integral transforms}

We can think of the covariance function as a kernel of an linear integral transform that defines an operator
\begin{align}
\hat{T}_k f(\vec{x}) = \int _{\mathcal{X} }k(\vec{x},\vec{x}')f(\vec{x}') \, {\mathrm {d}}\vec{x}' \text{,}
\label{eq:integral_operator}
\end{align}

which maps a function $f(\vec{x})$ from one domain to another~\cite{rasmussen2004gaussian, smola1998connection}. In this view, $\hat{T}_k f(\vec{x}) = \hat{f}(\vec{x})$ corresponds to the posterior mean of our GP. Note, that $\hat{T}_k f(\vec{x})$ remains a continuous function, even if we discretize the integration domain. This is the case in the regression setting, when we are only able to observe our target function partially, i.e. when $f(\vec{x}') = \sum_{i=1}^M{f(\vec{x}_i)\delta(\vec{x}'-\vec{x}_i)}$. With that in mind, an integral operator can be regarded as a continuous generalization of the matrix-vector product using a square matrix with entries $(\mat{K})_{ij} = k(\vec{x}_i, \vec{x}_j)$ and a vector $\vec{\alpha}$. Then,
\begin{align}
(\mat{K}\vec{\alpha})_i = \sum^M_j k(\vec{x}_i, \vec{x}_j) \alpha_j
\end{align}
is the discrete analogon to $\hat{T}_k f(\vec{x})$~\cite{heil2018metrics}. Note, that this expression is closed under linear transformation: any linear constraint $\hat{G}[\hat{T}_k]$ propagates into the integral and gives rise to a new covariance function.

However, there are several alternative construction options, one of them through explicit definition of the frequency spectrum of $\hat{T}_k$. Due to the translational symmetry of physical systems, we are particularly interested in stationary covariance functions that only depend on pairwise distances $\vec{\delta} = \vec{x} - \vec{x}'$ between points. In that setting, \emph{Bochner's theorem} says that symmetric, positive definite kernels can be constructed via the inverse Fourier transform of a probability density function $p(\vec{\delta})$ in frequency space~\cite{mackay1998introduction,smola1998connection,rasmussen2004gaussian,rahimi2008random}:
\begin{align}
k(\vec{\delta})=\mathcal{F}(p(\vec{\delta}))=\int p(\omega) e^{-i \omega\tran \vec{\delta}} \mathrm{d} \omega \text{.}
\end{align}

The following perspective might however be more intuitive when approaching this problem form a physics background: Since $\hat{T}_k f(\vec{x})$ is the reconstruction from point-wise observations $y_i = f(\vec{x}_i)$, we are ideally looking for an operator that leaves our unknown target function invariant, such that $\hat{T}_k f(\vec{x}) = f(\vec{x})$. This is another way of saying that our estimate $\hat{f}(\vec{x})$ lives in the space spanned by the eigenfunctions $\varphi_i \in \mathcal{F}$ of the operator defined by the kernel function (with coefficients $c_i \in \mathbb{R}$), giving
\begin{align}
\hat{f}(\vec{x}) = \sum_i c_i \varphi_i(\vec{x}) \quad \text{ with } \quad \hat{T}_k \varphi_i = \lambda_i \varphi_i \text{.}
\end{align}

It is impossible to overlook that there is a strong analogy between the covariance function in a GP process and the Hamiltonian in the Schr\"odinger equation (SE). Both operators formulate constraints that give rise to Hilbert space of possible states of the modeled object, whether it is the wavefunction or the hypothesis space of the GP. Although this is where the similarities end, this connection certainly illustrates that GPs are particularly suitable to reconstruct physical processes in a principled way.

\subsection{Mean function}

In most applications, the GP prior mean function $\mu(\vec{x}) = 0$ is set to zero, which leads to predictions $\hat{f}(\vec{x}) \approx 0$ as $\|\vec{x} - \vec{x}'\| \rightarrow 0$ for stationary kernels. Convergence to a constant outside of the training regime is desirable for data-driven models, because it means that the prediction degrades gracefully in the limit, instead of producing unforeseeable results. However, if a certain \emph{asymptotic} behavior of the modeled function is known, the prior mean function offers the possibility to prescribe it. For example, we could introduce a log barrier function
\begin{align}
\mu(\vec{x}) = -\log(\vec{b}-\vec{x})
\end{align}
that ramps up the predicted quantity towards infinity for $\vec{x} \geq 0$. In a molecular PES model, such a barrier would represent an atom dissociation limit, which could be useful to ensure that a dynamical process stays confined to the data regime as it moves around the PES.

In the spirit of how the Slater determinant in quantum mechanics accounts for the average affect of electron repulsion without explicit correlation, the mean of a GP can be used to prescribe a sensible predictor response outside of the data regime.

\section{Energy-conserving force field reconstructions}

%Noether's theorem~\cite{noether1918invarianten} states that each conserved quantity is associated with a differentiable symmetry of the action of a physical system. The action is represented as the integral 
%
%\begin{align}
%{\mathcal  {S}}=\int _{{t_{1}}}^{{t_{2}}}L \left[\vec{q}(t),\dot{\vec{q}}(t), t \right]\,\mathrm{d}t \text{,}
%\end{align}
%
%of the Lagrangian $L$ over time, were $\dot{\vec{q}} = \partial \vec{q} / \partial t$ is the change in the coordinates $\vec{q}$ over time. The behavior of any dynamical system is then described by the trajectories through this phase space for which the action is stationary. Consequently, a symmetry of a system is defined as any coordinate $q_k$ that does not appear on the Lagrangian, with the results that ${\partial L}/{\partial q_k} = 0$. Then, we have for the Euler-Lagrange equation of motion 
%
%\begin{align}
%\frac{\partial}{\partial t} \left( \frac{\partial L}{\partial \dot{q}_k} \right) = \frac{\partial L}{\partial q_k} = 0 \qquad \rightarrow \qquad \frac{\partial L}{\partial \dot{q}_k} = C \text{,}
%\end{align}
%
%where $C$ is a constant and ${\partial L}/{\partial \dot{q}_k}$ is a conserved quantity, i.e. independent of time. Conserved quantities in Lagrangian systems include the angular and linear momentum (roto-translational invariance), which can be easily implemented via   

%Conserved quantities in Lagrangian systems include the total energy (following from temporal invariance), as well as angular and linear momentum (roto-translational invariance).

A fundamental property that any molecular force field ${\mathbf F}(\vec{r}_1,\vec{r}_2,\dotsc,\vec{r}_N)$ must satisfy is the conservation of total energy, which implies that 
\begin{align}
{\mathbf F}(\vec{r}_1,\vec{r}_2,\dotsc,\vec{r}_N) = -\nabla E(\vec{r}_1,\vec{r}_2,\dotsc,\vec{r}_N)\text{.} 
\end{align}
While any analytically derivable expression for the potential energy satisfies energy conservation by definition, a direct reconstruction of a force field is more involved. It requires mapping to an explicitly conservative vector field and thus special constraints on the hypothesis space that the ML model navigates in. Before we discuss how this can be implemented in practice, we will briefly review a number of advantages that make this direct approach worthwhile.

\subsubsection{Forces are quantum-mechanical observables}

A major reason is the fact that atomic forces are true quantum-mechanical observables within the BO approximation by virtue of the \emph{Hellmann-Feynman theorem}. It provides a way to obtain analytical derivatives by relating changes in the total energy $\delta E$ with respect to any variation $\delta \lambda$ of the Hamiltonian $H$ through the expectation value
\begin{align}
{\frac  {\partial E}{\partial \lambda}} = {\bigg \langle }\Psi_\lambda {\bigg |}{\frac  {\partial {\hat {H}}_\lambda}{\partial \lambda}}{\bigg |}\Psi_\lambda {\bigg \rangle } \ \text{.}
\label{eq:hf_theorem}
\end{align}
It thus allows the direct computation of forces $\mat{F} =-  {\partial E} / {\partial \mat{R}}$ as derivatives with respect to nuclei positions $\mat{R}$. Similar analogous expressions for density based approaches exist as well~\cite{politzer2018hellmann, feynman1939forces}. Once the SE is solved for a particular atomic configuration to compute the energy, this theorem makes the additional computation of forces relatively cheap, by reusing some of the results (most importantly, the parametrization of the wave-function $\Psi$).

The appealing aspect about (analytic) force observations is that they are considerably more informative, as they represents a linearization of the PES in all directions of the $3N$-dimensional configuration space, instead of a single point evaluation. Gathering a similar amount of insight about the PES numerically via energy examples, would require solutions of the SE for at least $3N+1$ perturbations $E(r_1, \dots, r_i + \epsilon ,\dots, r_{3N})$ of the original geometry at each point. Even then, the obtained force would be subject to approximation error. In contrast, computing analytical forces using the Hellman-Feynman theorem only requires about one to seven times the computational effort of a single energy calculation. Effectively, this theorem thus offers a more efficient way to sample PESs. 

%provided that these can also be directly learned by the ML model.

%and likely inconsistent with the corresponding energy, when re-integrated

\begin{figure}[t]
\centering
\includegraphics[width=\textwidth]{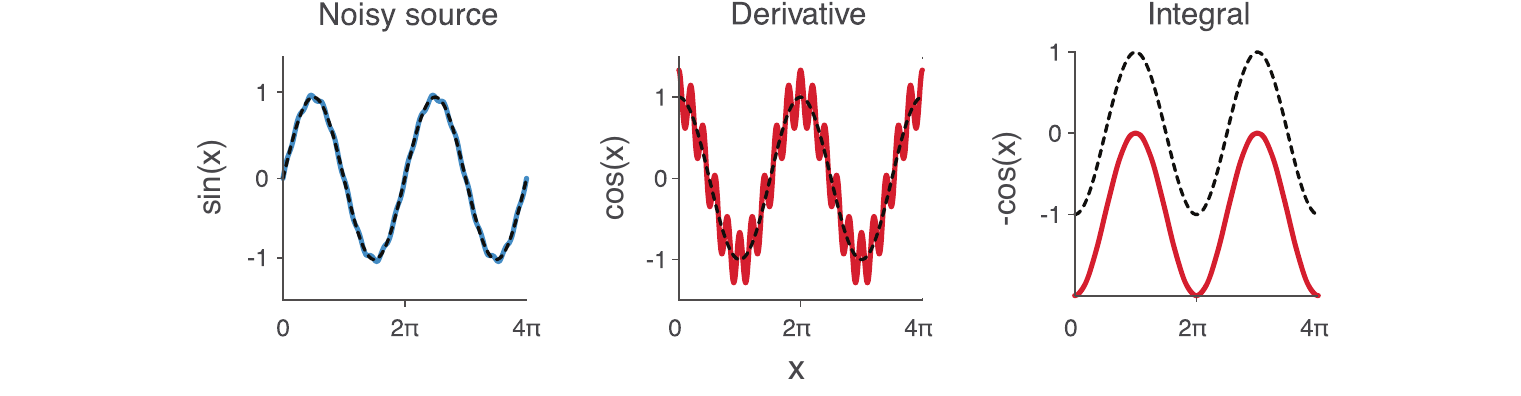}
\caption{A noisy approximation of a sine wave (blue). Although all instantaneous values are represented well, the derivative of the approximation is a poor estimator for the true derivative. This is because differentiation amplifies the high frequency noise component within the approximation (middle). Integration on the other hand acts as a low-pass filter (right) that attenuates noise. It is therefore easier to approximate a function with accurate first derivatives from derivative examples instead of function values. Note that integrals are only defined up to an additive constant, which needs to be recovered separately. Figure taken from Chmiela, 2019 \cite{chmiela2019thesis}.
}
\label{fig:noisy_derivatives}
\end{figure}

\subsubsection{Differentiation amplifies noise}

When force estimates are obtained via differentiation of an \emph{approximate} PES reconstruction, there are no guarantees regarding their quality, since forces are neither constrained nor directly regularized within the loss function of an energy-based model. Inevitably, this can lead to artifacts.

Reconstructions of functions based on a limited number of observations will almost always not be error-free, either due to aliasing effects, non-ideal choice of hypothesis space or noisy training data~\cite{scholkopf2001generalized}. Furthermore, the regularization term in the loss function will reduce the variance of the model and thus promote any approximation errors into the high-frequency band of the residual $f - \hat{f}$. Unfortunately, the application of the derivative operator amplifies high frequencies $\omega$ with increasing gain~\cite{Shannon1998}, drastically magnifying these errors. This phenomenon can be easily understood when looking at the derivative of a model $\hat{f}'$ in the frequency domain
\begin{equation}
\mathcal{F}\left[ \hat{f}' \right] = \mathrm{i} \omega \mathcal{F}\left[\hat{f}\right] \text{,}
\end{equation}
were $\omega$ is a factor on the Fourier transform $\mathcal{F}[\hat{f}]$ of the original model (see Figure~\ref{fig:noisy_derivatives}). A low test error in the energy prediction task does therefore not necessarily imply that the model also reconstructs the forces of the target function reliably. Conversely, however, high-frequency noise is attenuated upon integration of a force field estimate, which gives corresponding potentials with controlled variance.

\begin{figure}[t]
\centering
\includegraphics[width=\textwidth]{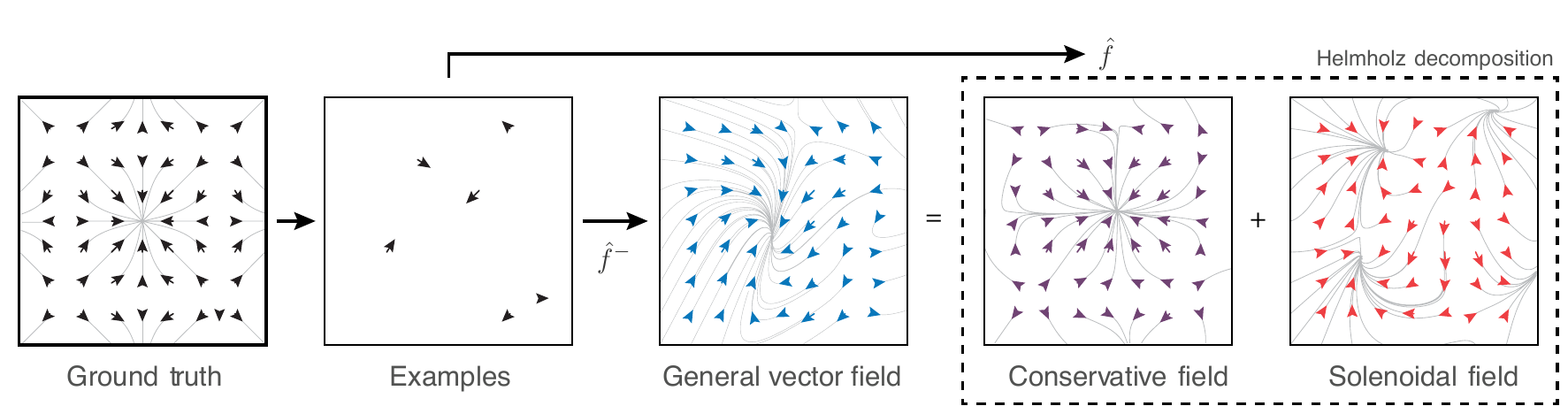}
\caption{Modeling gradient fields (leftmost subfigure) based on a small number of examples. With GDML, a conservative vector field estimate $\mat{\hat{f}}$ is obtained directly (purple). In contrast, a na\"{\i}ve estimator $\mat{\hat{f}^-}$ with no information about the correlation structure of its outputs is not capable to uphold the energy conservation constraint (blue). We perform a Helmholtz decomposition of the na\"{\i}ve non-conservative vector field estimate to show the error component due violation of the law of energy conservation (red). This significant contribution to the overall prediction error is completely avoided with the GDML approach. Figure taken from Chmiela et al., 2017}
\label{fig:helmholtz_decomp}
\end{figure}

\subsection{Constructing conservative vector-valued GPs}

In the simplest, and by far most prevalent regression setting, a \emph{single} output variable $y \in \mathbb{R}$ is predicted from an input vector $\vec{x} \in \mathbb{R}^N$. Being a scalar field, the PES reconstruction problem fits this template, however we intent to pursue the reconstruction of the associated force field, i.e. the negative gradient of the PES, instead. In that reformulation of the problem, the output $\vec{y} \in \mathbb{R}^N$ becomes vector-valued as well, thus requiring a mapping $\mat{\hat{f}}: \mathbb{R}^N \rightarrow \mathbb{R}^N$.

Naively, we could model each partial force separately and treat them as independent, implicitly assuming that the individual components of the force vector do not affect each other. Then, a straightforward formulation of a vector-valued estimator takes the form
\begin{align}
\mat{\hat{f}}(\vec{x}) = [\hat{f}_1(\vec{x}), \dots, \hat{f}_N(\vec{x})]\tran \text{,}
\end{align}
where each component $\hat{f}_i: \mathbb{R}^N \rightarrow \mathbb{R}$ is a separate scalar-valued GP~\cite{friedman2001elements}. However, this assumed independence of the individual outputs is hard to justify in many practical scenarios. Especially, since correlations between the individual noise processes associated with each output channel could introduce dependencies in the posterior process, even if they were independent \emph{a priori}~\cite{rasmussen2004gaussian}. Bypassing this dependence would therefore ignore valuable information and yield sub-optimal estimates.

%An artificial decoupling would therefore ignore valuable information and yield sub-optimal estimates.

%This would be an unfortunate conclusion, because multivariate output dependencies can be naturally captured by GPs through the correlation structure of the prior.

Instead of mapping to scalar outputs, we can alternatively model the covariance function as a matrix $\mat{k}: \mathcal{X} \times \mathcal{X} \rightarrow \mathbb{R}^{N \times N}$ that expresses the interaction among multiple output components. Together with a vector-valued mean function $\mat{\mat{\mu}}: \mathcal{X} \rightarrow \mathbb{R}^{N}$, we can then sample realizations of vector-valued functions from the GP
\begin{align}
\mat{f}(\vec{x}) \sim \mathcal{GP} \left[\mat{\mu}(\vec{x}), \mat{k}(\vec{x},\vec{x}') \right] \text{.}
\end{align}
In this setting, the corresponding RKHS is vector-valued and it has been shown that the representer theorem continues to hold~\cite{alvarez2012kernels}.
Each component of the kernel function $k_{ij}$ specifies a covariance between a pair of outputs $f_i(\vec{x})$ and $f_j(\vec{x})$, which makes it straightforward to impose linear constraints $\mat{g}(x) = \hat{G}\left[\mat{f}(\vec{x})\right]$ on the GP prior
\begin{align}
\mat{g}(\vec{x}) \sim \mathcal{GP} \left[\hat{G}\mat{\mu}(\vec{x}), \hat{G}\,\mat{k}(\vec{x},\vec{x}')\,\hat{G'}\tran \right] \text{.}
\end{align}
and hence also the posterior~\cite{boyle2005dependent,micchelli2005learning,micchelli2005kernels,baldassarre2012multi}. Here, $\hat{G}$ and $\hat{G'}$ act on the first and second argument of the kernel function, respectively. Linear constraints are wide-spread in physics. They include simple conservation laws, but also operations like differential equations, allowing the construction of models that are consistent with the laws that underpin many physical processes~\cite{graepel2003solving,sarkka2011linear,constantinescu2013physics,nguyen2015gaussian,jidling2017linearly}. 

%Single output GPs are included as a special case in this multiple output generalization: setting the matrix-valued kernel function $\mat{k}(\vec{x}, \vec{x}') = k(\vec{x}, \vec{x}') \mat{\mathbb{1}}_{N}$ to a diagonal matrix treats all outputs as independent and hence recovers the decoupled-output GP.

Here, we aim to construct a GP that inherits the correct structure of a \emph{conservative} force field to ensure integrability, so that the corresponding energy potential can be recovered from the same model. We start by considering, that the force field estimator $\mat{\hat{f}_F}(\vec{x})$ and the PES estimator $\hat{f}_E(\vec{x})$ are related via some operator $\hat{G}$. To impose energy conservation, we require that the curl vanishes (see Figure~\ref{fig:helmholtz_decomp}) for every input to the transformed energy model\footnote{For illustrative purposes, we use the definition of curl in three dimensions here, but the theory directly generalizes to arbitrary dimension. One way to prove this is via path-independence of conservative vector fields: the circulation of a gradient along any closed curve is zero and the curl is the limit of such circulations.}:  
\begin{align}
\nabla \times \hat{G}\left[\hat{f}_E\right] = \vec{0} \text{.}
\end{align}
As expected, this is satisfied by the derivative operator $\hat{G} = \nabla$ or, in the case of energies and forces, the negative gradient operator 
\begin{align}
\mat{\hat{f}_F}(\vec{x}) = \hat{G}\left[\hat{f}_E\right](\vec{x}) = -\nabla \hat{f}_E(\vec{x}) \text{.}
\end{align}
As outlined previously, we can directly apply this transformation to a standard scalar-valued 'energy' GP with realizations $f_E: \mathcal{X}^{3N} \rightarrow \mathbb{R}$. Since differentiation is a linear operator, the result is another GP with realizations $\mat{f_F}: \mathcal{X}^{3N} \rightarrow \mathbb{R}^{3N}$:
\begin{align}
\mat{\hat{f}_F} \sim \mathcal{GP} \left[-\nabla\mu(\vec{x}), \nabla_{\vec{x}} k(\vec{x},\vec{x}') \nabla\tran_{\vec{x}'}\right] \text{.}
\label{eq:derivative_gp}
\end{align}

Note, that this gives the second derivative of the original kernel (with respect to each of the two inputs) as the (co-)variance structure, with entries
\begin{align}
k_{ij} = \frac{\partial^2 k}{\partial \vec{x}_i \partial \vec{x}'_j} \text{.}
\end{align}
It is equivalent (up to sign) to the Hessian $\nabla k \nabla\tran = \text{Hess}_{\vec{x}}(k)$ (i.e. second derivative with respect to one of the inputs), provided that the original covariance function $k$ is stationary. A GP using this covariance enables inference based on the distribution of partial derivative observations, instead of function values~\cite{narcowich1994generalized,solak2003derivative}. Effectively, this allows us to train GP models in the gradient domain.

This Hessian kernel gives rise to the following \emph{Gradient Domain Machine Learning}~\cite{gdml,chmiela2019} force model as the posterior mean of the corresponding GP:
\begin{equation}
  \begin{aligned}
  \mat{\hat{f}_F} (\vec{x}) = \sum^M_i \sum^{3N}_j  (\vec{\alpha_i})_j  \frac{\partial}{\partial x_j} \nabla k(\vec{x}, \vec{x}_i)
  \label{eq:force_model_gp_mean}
    \end{aligned}
\end{equation}
Because the trained model is a (fixed) linear combination of kernel functions, integration only affects the kernel function itself. The corresponding expression for the energy predictor
\begin{equation}
  \begin{aligned}
  \hat{f}_E (\vec{x}) = \sum^M_i \sum^{3N}_j  (\vec{\alpha_i})_j  \frac{\partial}{\partial x_j} k(\vec{x}, \vec{x}_i) + c
  \label{eq:gdml_energy_model}
    \end{aligned}
\end{equation}
is therefore neither problem-specific, nor does it require retraining. It is however only defined up to an integration constant
\begin{equation}
c = \frac{1}{M}{\sum^M_i  E_i + \hat{f}_E(\vec{x}_i)} \text{,}
\end{equation}
that we recover separately in the least-squares sense. Here, $E_i$ are the energy labels for each training example. We remark that this reconstruction approach yields two models at the same time, by correctly implementing the fundamental physical connection between $\mat{k_F}$ and $k_E$.

\begin{figure}[t]
  \begin{minipage}{\columnwidth} 
     \centering 
     \includegraphics[width=0.7\columnwidth]{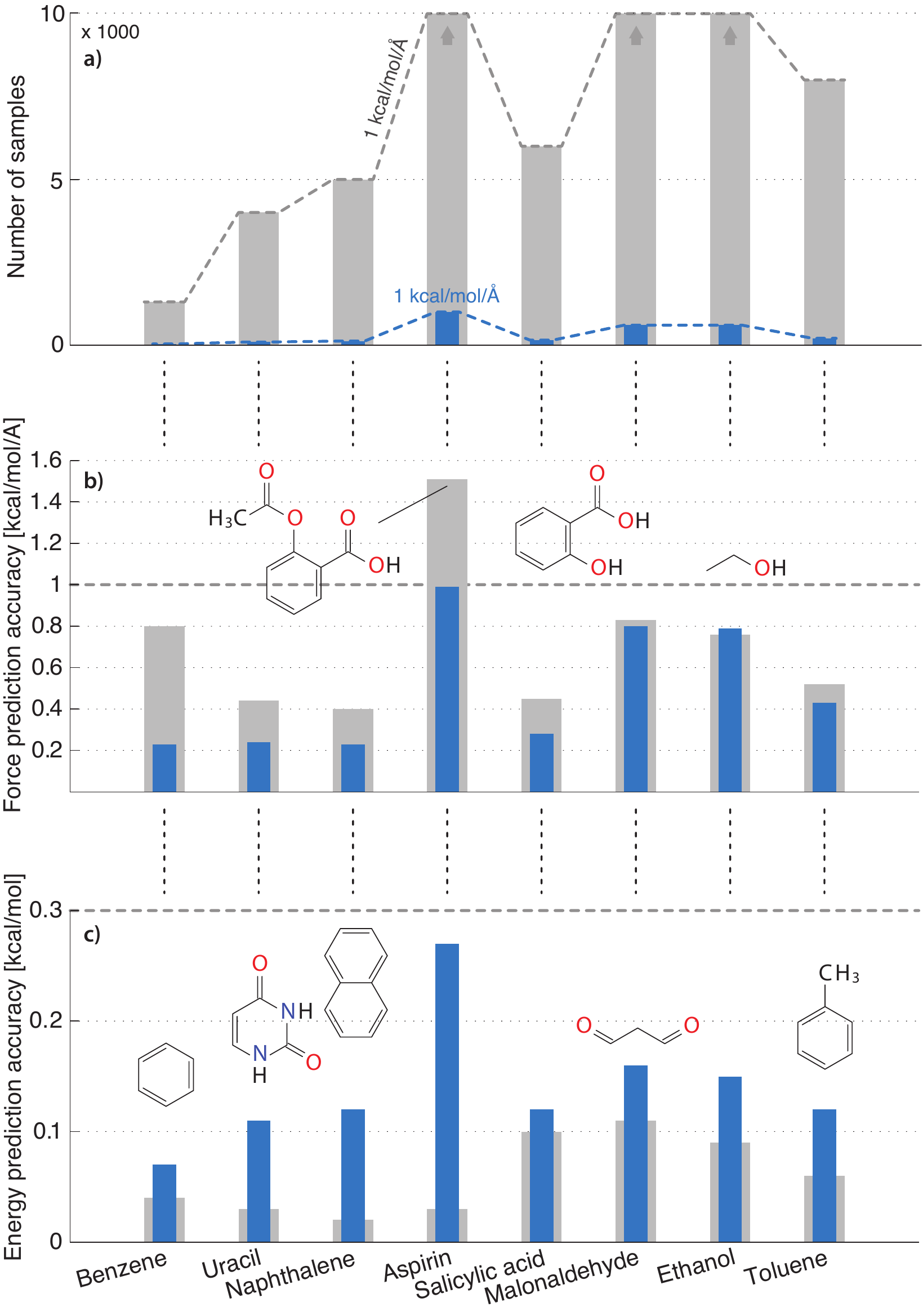} 
  \end{minipage} 
\caption{Efficiency of GDML predictor versus a model that has been trained on energies. (a) Required number of samples for a force prediction performance of MAE (1 kcal $\text{mol}^{-1}$ $\text{\AA}^{-1}$) with the energy-based model (gray) and GDML (blue). The
energy-based model was not able to achieve the targeted performance with the maximum number of 63,000 samples for aspirin. (b) Force prediction errors for the converged models (same number of partial derivative samples and energy samples). (c) Energy prediction errors for the converged models. All reported prediction errors have been estimated via cross-validation. Figure taken from Chmiela et al., 2017 \cite{gdml}.}
\label{fig:bars}
\end{figure}

\section{Point groups and fluxional symmetries}

%Molecules possess well-defined rigid space group symmetries (e.g. reflections), as well as dynamic non-rigid symmetries (e.g. rotations of functional groups).

%These can be expressed as permutations of atomic indices within the numerical representation of the molecule that leave the corresponding energy and forces invariant.

Atoms of the same species are interchangeable within a numerical representation of the molecule, without affecting the corresponding energy or forces. This permutational invariance is directly inherited from the indistinguishability of identical atoms in the nuclear Hamiltonian 
\begin{equation}
H_{n}=T_n + E_{tot}=T_n + E_{nn}+E_{e} \text{.} 
\end{equation}
Here, $T_n$ and $E_{nn}$ are the kinetic an electrostatic energies if the nuclei and $E_{e}$ is the eigenvalue of the electronic Hamiltonian $H_{e}$. $E_{e}$ inherits the parametrical dependence on the nuclear coordinates of $H_{e}$, which is invariant with respect to atomic permutations given its analytical form: $H_{e}=T_{e}+E_{ee}+\sum_{i=1}^{N_n}{\sum_{j=1}^{N_e}{\frac{1}{|\mat{R}_i-\mat{r}_j|}}}$, where the $T_e$ and $E_{ee}$ are the electronic kinetic an electrostatic energies. Since the dependence of $H_{e}$ on the nuclear coordinates appears only as a sum, we can freely permute the nuclear positions, defining the permutations symmetry over the whole symmetric group.

For example, the benzene molecule with six carbon and six hydrogen atoms can be indexed (and therefore represented) in $6!6! = 518400$ different, but physically equivalent ways. However, not all of these symmetric variants are actually important to define a kernel function that gauges the similarity between pairs of configurations. To begin with, the 24 symmetry elements in the $\mat{D}_{6h}$ point group of this molecule are relevant. In addition to these rigid space group symmetries (e.g. reflections), there are additional dynamic non-rigid symmetries~\cite{Longuet-Higgins1963} (e.g. rotations of functional groups or torsional displacements) that appear as the structure transforms over time. Fortunately, we can safely ignore the massive amount of the remaining configurations in the full symmetric group of factorial size, if we manage to identify this relevant subset. While methods for identifying molecular point groups for polyatomic rigid molecules are readily available~\cite{wilson1955molecular}, dynamical symmetries are usually not incorporated in traditional force fields and electronic structure calculations. This is because extracting nonrigid symmetries requires chemical and physical intuition about the system at hand, which is hard to automate.

Since this is impractical in an ML setting, we will now review a physically motivated algorithm for purely data driven recovery of this particularly important subgroup of molecular symmetries. This will allow us to incorporate these symmetries as prior knowledge into a GP to further improve the data-efficiency of the model.

%To automate that process we will develop a physically motivated algorithm for data driven discovery of all relevant molecular symmetries from MD trajectories.

%Before we can incorporate these symmetries as prior knowledge into our GP, we develop a physically motivated algorithm for data driven discovery of all relevant molecular symmetries from MD trajectories. This will allow us to constrain the hypothesis space of the GP and further improve the data-efficiency of the model.

\subsection{Positive-semidefinite assignment}

We begin with the basic insight that MD trajectories consist of smooth consecutive changes in nearly isomorphic molecular graphs. When sampling from these trajectories the combinatorial challenge is to correctly identify the same atoms across the examples such that the learning method can use consistent information for comparing two molecular conformations in its kernel function. While so-called bi-partite matching allows to locally assign atoms $\mat{R} = \{\vec{r}_1, \dots, \vec{r}_N\}$ for each pair of molecules in the training set, this strategy alone is not sufficient as the assignment needs to be made globally consistent by multipartite matching in a second step~\cite{pachauri2013solving, Schiavinato2015, Kriege2016}. The reason is that optimal bi-partite assignment yields indefinite functions in general, which are problematic in combination with kernel methods. They give rise to indefinite kernel functions, which do not define a Hilbert space~\cite{Vert2008}. Practically, there will not exist a metric space embedding of the complete set of approximate pairwise similarities defined in the kernel matrix and the learning problem becomes ill-posed. A multipartite correction is therefore necessary to recover a non-contradictory notion of similarity across the whole training set. A side benefit of such a global matching approach is that it can robustly establish correspondence between distant transformations of a geometry using intermediate pairwise matchings, even if the direct bi-partite assignment is not unambiguously possible.

\subsubsection{Solving the multi-way matching problem}

We start by defining the bi-partite matching problem in terms of adjacency matrices as representation for the molecular graph. To solve the pairwise matching problem we therefore seek to find the assignment $\tau$ which minimizes the squared Euclidean distance between the adjacency matrices $\mat{A}$ of two isomorphic graphs $G$ and $H$ with entries $(\mat{A})_{ij} = \|\vec{r}_i - \vec{r}_j\|$, where $\mat{P}(\tau)$ is the permutation matrix that realizes the assignment:
\begin{equation}
\operatorname*{arg\,min}_{\tau} \mathcal{L}(\tau) = \|\mat{P}(\tau)\mat{A}_G\mat{P}(\tau)\tran - \mat{A}_H\|^2 \text{.}
\label{eq:matching_objective}
\end{equation}
Notably, most existing ML potentials use representations based on adjacency matrices as input~\cite{rupp2012fast,Hansen2013,Hansen2015,Rupp2015,Hirn2015,deltaML2015,Ceriotti2016,artrith2017efficient,Ceriotti2017,Glielmo2017,yao2017many,john2017many,faber2017prediction,eickenberg2018solid,glielmo2018efficient,tang2018atomistic,Grisafi2018,pronobis2018many,Behler2007,Bartok2010,Behler2012,Bartok2013,Montavon2013a,Bartok2015,Ramprasad2015,Tristan2015,DeVita2015,Behler2016,Brockherde2017,Gastegger2017,Schutt2017,schutt2017schnet,SchNet2018,huang1707dna,huan2017universal,Shapeev2017,dral2017structure,zhang2018deep,lubbers2018hierarchical,ryczko2018convolutional,kanamori2018exploring,hy2018predicting,Clementi2018,Tristan,noe2018,noe2018b,thomas2018tensor,FCHL2018,christensen2019operators,unke2019physnet,winter2019}. An optimal assignment in terms of Eq.~\ref{eq:matching_objective} therefore transfers to almost any other model and the GDML model in particular.

Adjacency matrices of isomorphic graphs have identical eigenvalues and eigenvectors, only their assignment differs. Following the approach of
Umeyama~\cite{Umeyama1988}, we identify the correspondence of eigenvectors $\mat{U}$ by projecting both sets $\mat{U}_G$ and
$\mat{U}_H$ onto each other to find the best overlap. We use the overlap matrix,
\begin{equation}
\mat{M} = \text{abs}(\mat{U}_G) \text{abs}(\mat{U}_H)\tran
\label{eq:cost_matrix}
\end{equation}
after sorting eigenvalues and overcoming sign ambiguity. Then $-\mat{M}$ is provided as the cost matrix for the Hungarian algorithm~\cite{Kuhn1955}, maximizing the overall overlap which finally returns the approximate assignment $\tilde{\tau}$ that minimizes Eq.~\ref{eq:matching_objective} and thus provides the results of step one of the procedure. As indicated, global inconsistencies may arise, observable as violations of the transitivity property $\tau_{jk} \circ \tau_{ij} =\tau_{ik}$ of the assignments~\cite{pachauri2013solving}. Therefore a second step is necessary which is based on the composite matrix $\tilde{\mathcal{P}}$ of all pairwise assignment matrices $\tilde{\mat{P}}_{ij} \equiv \mat{P}(\tilde{\tau}_{ij})$ within the training set.
\begin{figure}[t]
\centering
\includegraphics[width=0.8\textwidth]{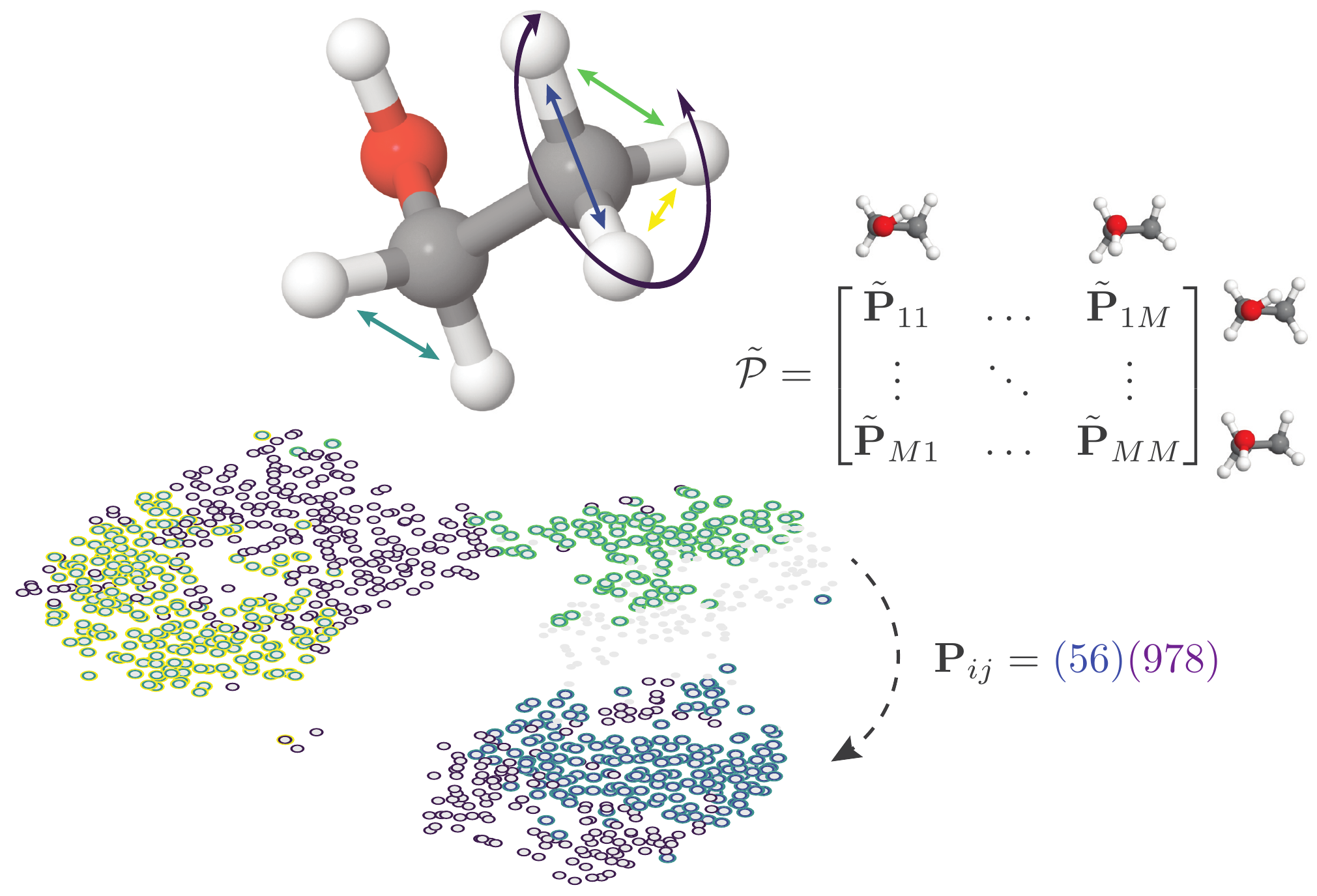}
\caption{T-SNE~\cite{vanderMaaten2008} embedding of all molecular geometries in an ethanol training set. Each data point is color coded to show the permutation transformations that align it with the arbitrarily chosen canonical reference state (gray points). These permutations are recovered by restricting the rank of the pairwise assignment matrix $\tilde{\mathcal{P}}$ to obtain a consistent multi-partite matching $\mathcal{P}$.}
\label{fig:recovered_syms}
\end{figure}
We propose to reconstruct a rank-limited $\mathcal{P}$ via the transitive closure of the minimum spanning tree (MST) that minimizes the bi-partite matching cost (see Eq.~\ref{eq:matching_objective}) over the training set. The MST is constructed from the most confident bi-partite assignments and represents the rank $N$ skeleton of $\tilde{\mathcal{P}}$, defining also $\mathcal{P}$ (see Figure~\ref{fig:recovered_syms}). Finally, the resulting \emph{multi-partite matching} $\mathcal{P}$ is a consistent set of atom assignments across the whole training set.

As a first test, we apply our algorithm to a diverse set of non-rigid molecules that have been selected by Longuet-Higgins~\cite{Longuet-Higgins1963} to illustrate the concept of dynamic symmetries. Each of the chosen examples changes easily from one conformation to another due to internal rotations that can not be described by point groups. Those molecules require the more complete \emph{permutation-inversion group} of symmetry operations that include energetically feasible permutations of identical nuclei. Our multi-partite matching algorithm successfully recovers those symmetries from short MD trajectories (see Table~\ref{tab:symmetry_recovery}), giving us the confidence to proceed.

\begin{table}[t]
\centering
\caption{Recovering the permutation-inversion (PI) group of symmetry operations of fluxional molecules from short MD trajectories. We used our multi-partite matching algorithm to recover the symmetries of the molecules used in Longuet-Higgins~\cite{Longuet-Higgins1963}. Our algorithm identifies PI group symmetries (a superset that also includes the PG), as well as additional symmetries that are an artifact of the metric used to compare molecular graphs in our matching algorithm. Each dataset consists of a MD trajectory of 5000 time steps. Figure taken from Chmiela, 2019 \cite{chmiela2019thesis}.}
\label{tab:symmetry_recovery}
\begin{tabular}{p{13mm} l r r r}
\toprule
~~~~~~~~~~~~ & \multicolumn{1}{l}{\textbf{Molecule}} & \multicolumn{1}{l}{\textbf{PG order}} & \multicolumn{1}{l}{\textbf{PI group order}} & \multicolumn{1}{l}{\textbf{Recovered}}\\
\hline
\parbox[c]{0em}{\includegraphics[scale=0.02]{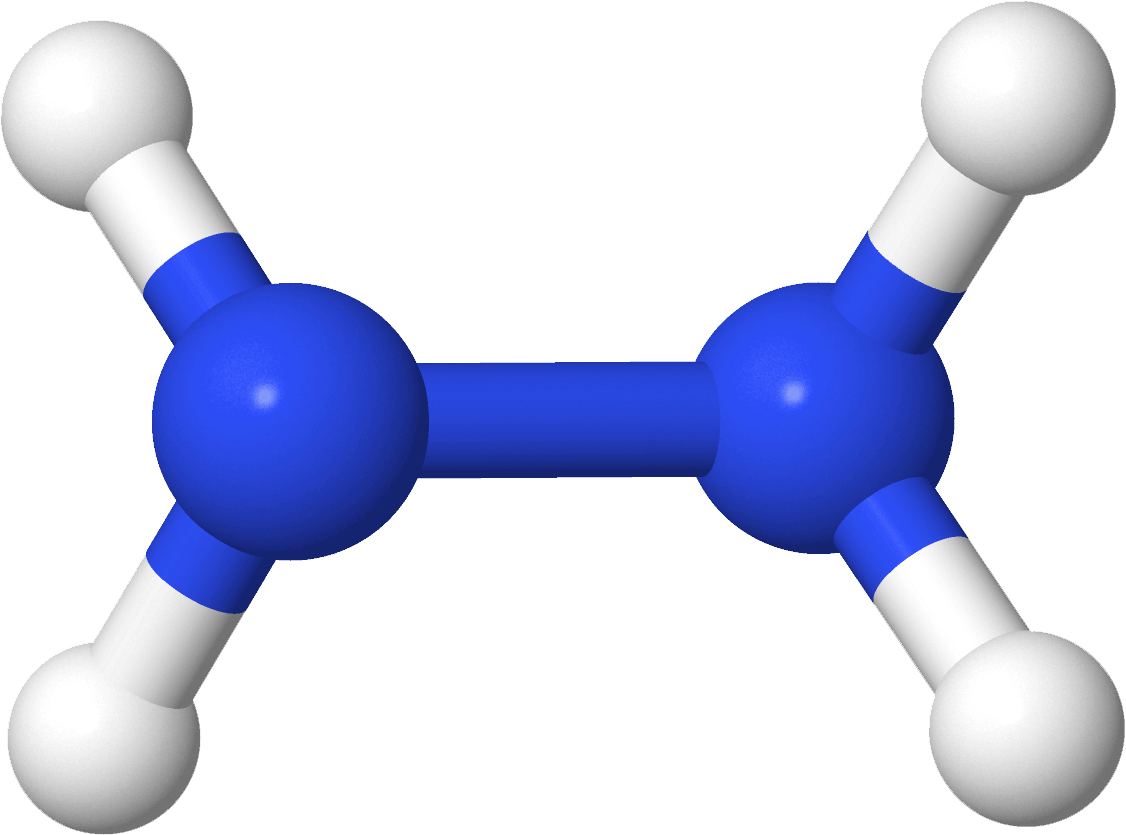}} &
Hydrazine &
\ensuremath{2} &
\ensuremath{8} &
\ensuremath{8}\\
\parbox[c]{0em}{\includegraphics[scale=0.02]{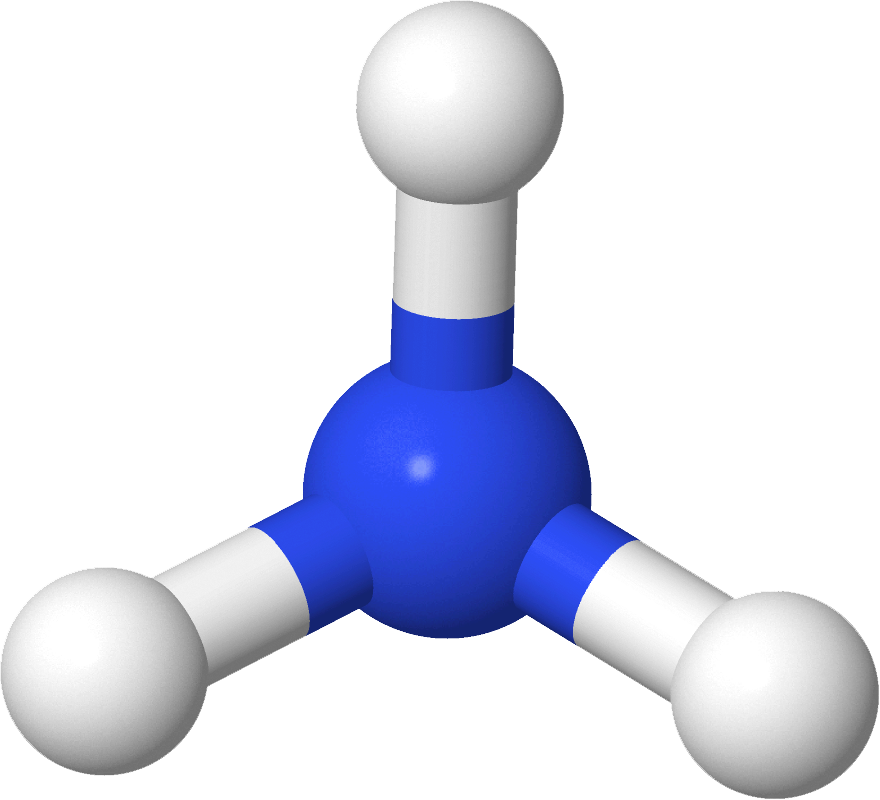}} &
Ammonia &
\ensuremath{6} &
\ensuremath{6} &
\ensuremath{6}\\
\parbox[c]{0em}{\includegraphics[scale=0.025]{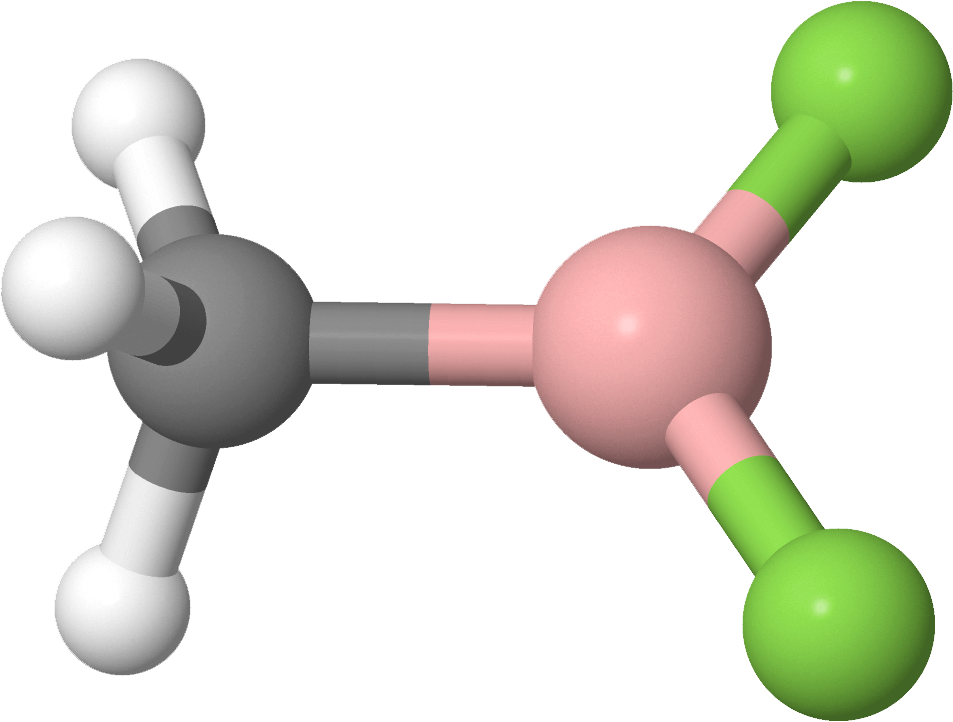}} &
(Difluoromethyl)borane &
\ensuremath{2} &
\ensuremath{12} &
\ensuremath{12}\\
\parbox[c]{0em}{\includegraphics[scale=0.025]{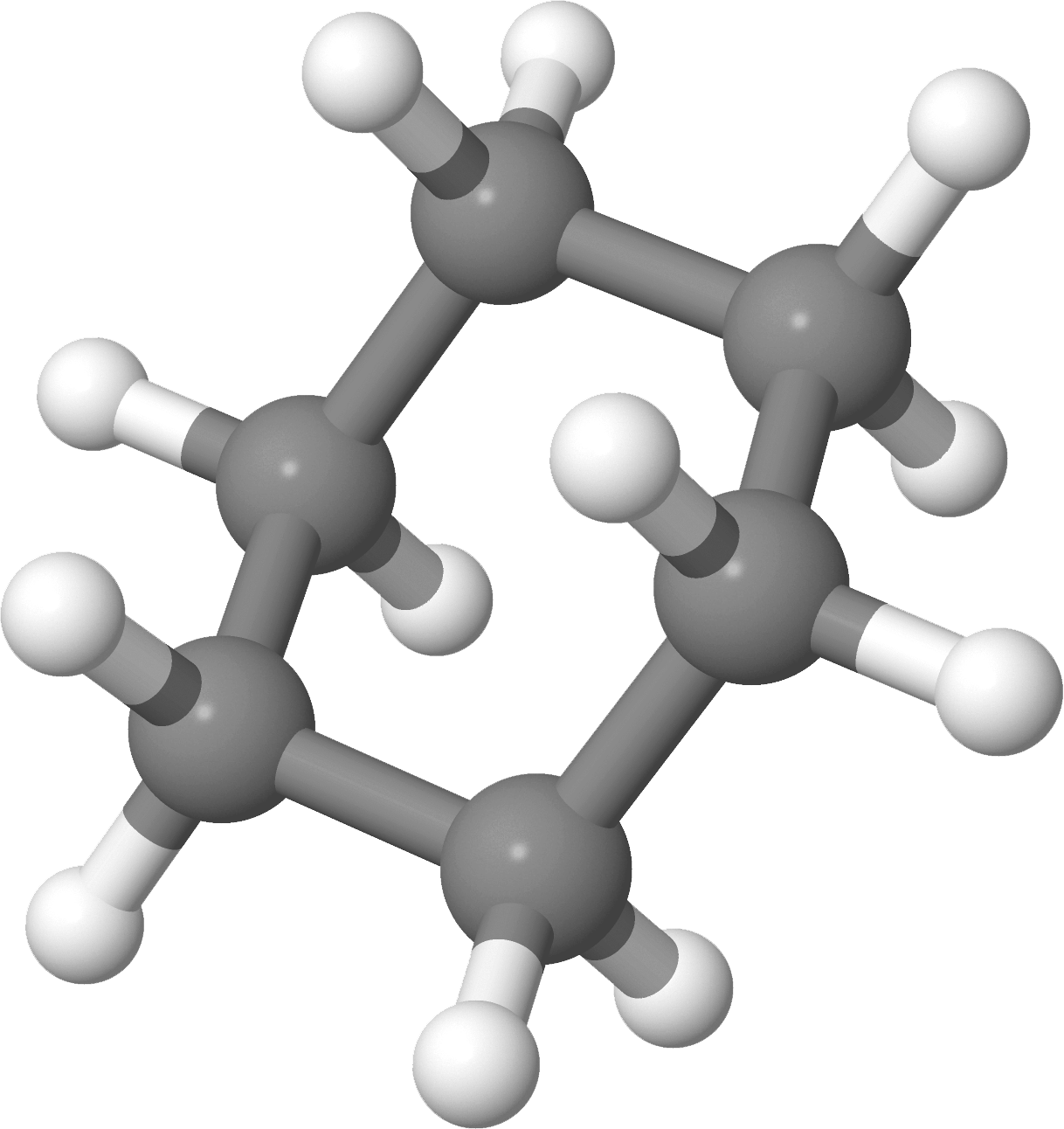}} &
Cyclohexane &
\ensuremath{6} &
\ensuremath{12} &
\ensuremath{12}\\
\parbox[c]{0em}{\includegraphics[scale=0.025]{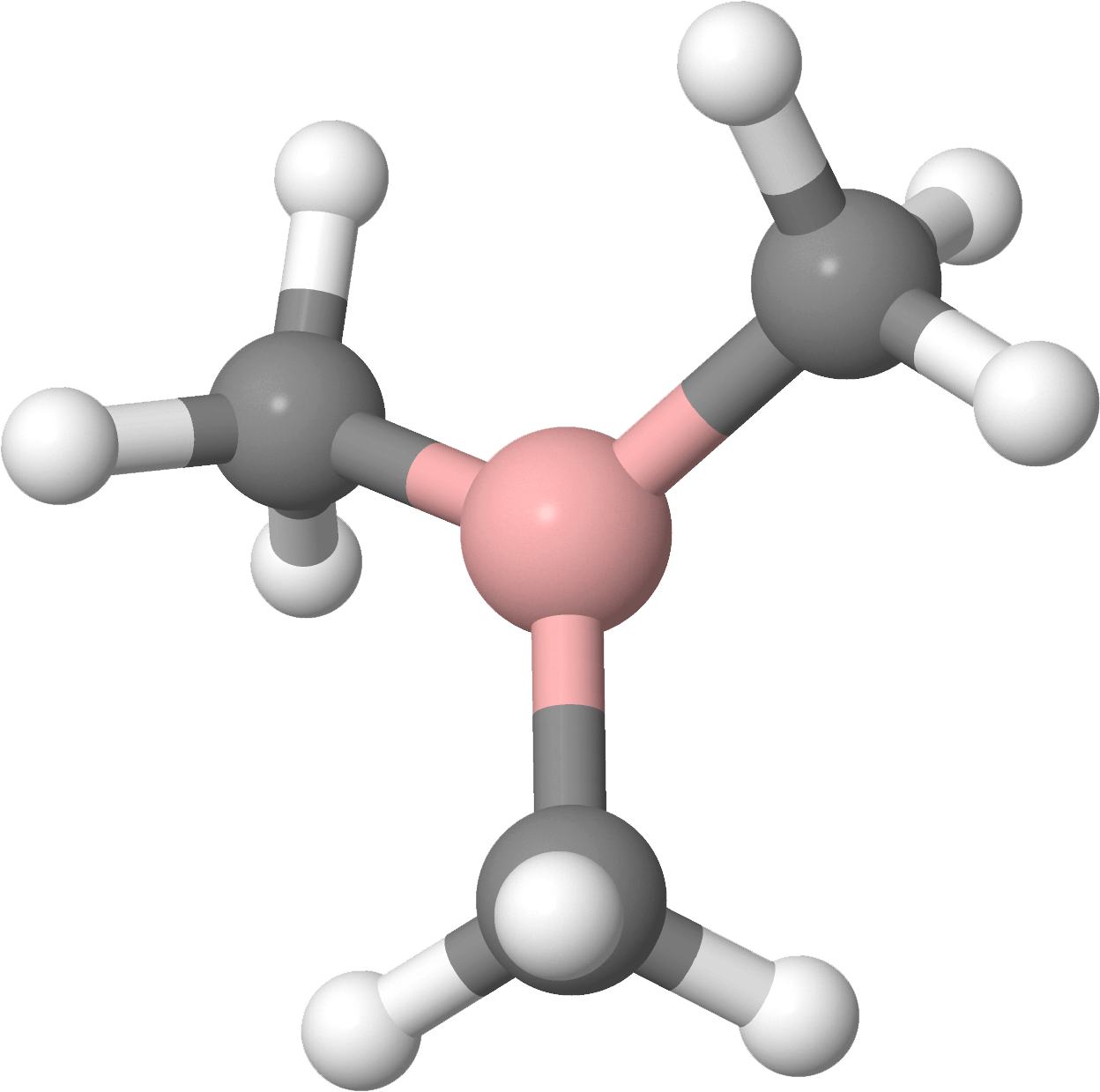}} &
Trimethylborane &
\ensuremath{2} &
\ensuremath{324} &
\ensuremath{339}\\
\parbox[c]{0em}{\includegraphics[scale=0.025]{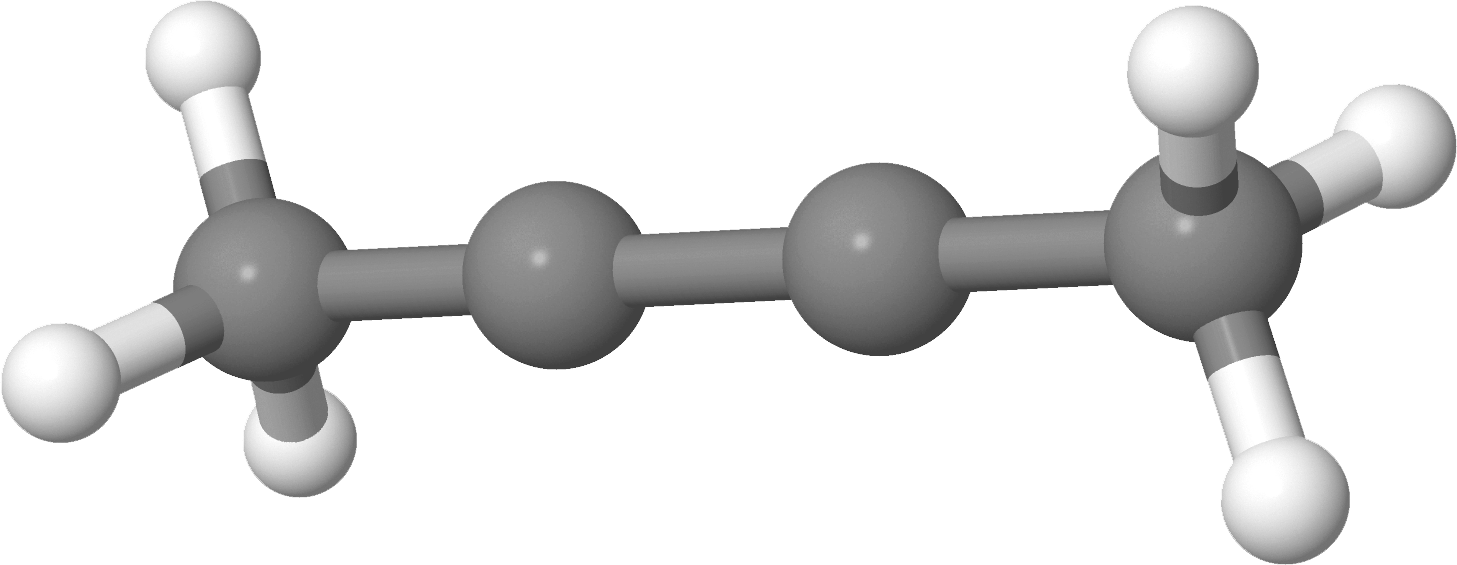}} &
Dimethylacetylene &
\ensuremath{6} &
\ensuremath{36} &
\ensuremath{39}\\
\parbox[c]{0em}{\includegraphics[scale=0.02]{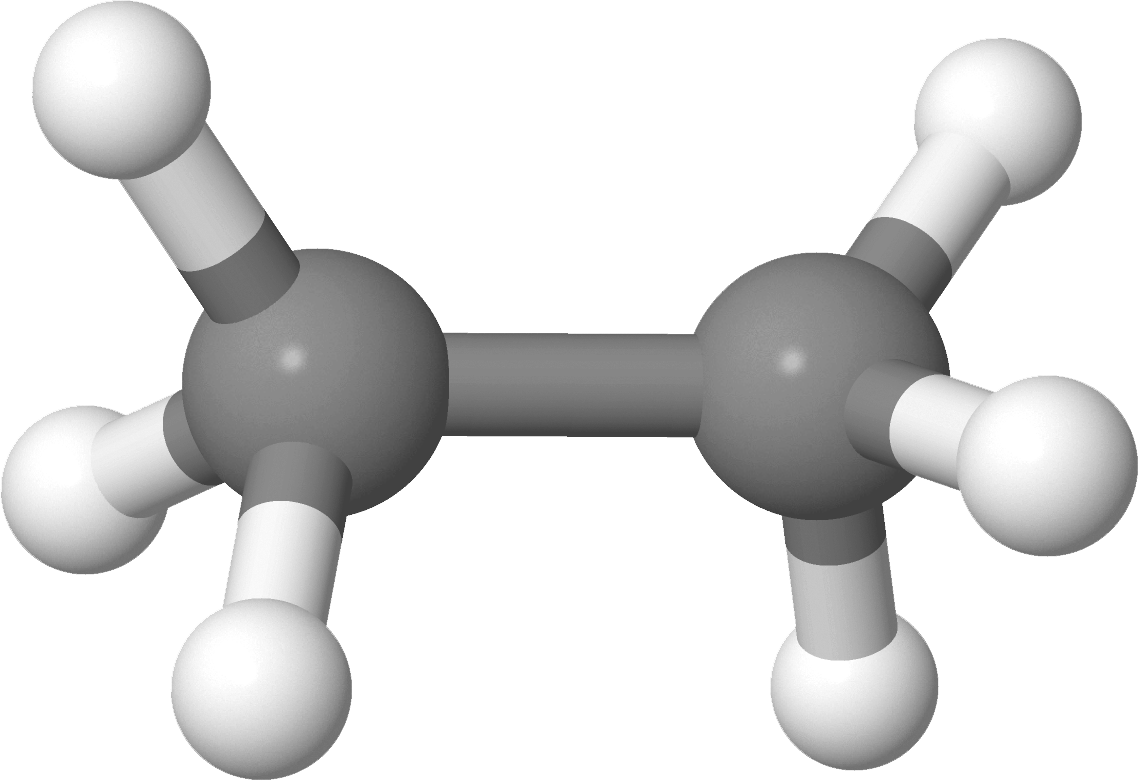}} &
Ethane &
\ensuremath{6} &
\ensuremath{16} &
\ensuremath{36}\\
    \bottomrule
\end{tabular}
\end{table}

\subsubsection{Symmetric kernels}

The resulting consistent multi-partite matching $\mathcal{P}$ enables us to construct symmetric kernel-based ML models of the form
\begin{equation}
\hat{f}(\vec{x}) = \sum^M_{ij} \alpha_{ij} k(\vec{x},\mat{P}_{ij} \vec{x}_i) \text{,}
\label{eq:data_augmentation_model}
\end{equation}
by augmenting the training set with the symmetric variations of each molecule~\cite{karvonen2018fully}. A particular advantage of our solution is that it can fully populate all recovered permutational configurations even if they do not form a symmetric group, severely reducing the computational effort in evaluating the model. Even if we limit the range of $j$ to include all $S$ unique assignments only, the major downside of this approach is that a multiplication of the training set size leads to a drastic increase in the complexity of the cubically scaling GP regression algorithm. We overcome this drawback by exploiting the fact that the set of coefficients $\vec{\alpha}$ for the symmetrized training set exhibits the same symmetries as the data, hence the linear system can be contracted to its original size, while still defining the full set of coefficients exactly.

Without affecting the pairwise similarities expressed by the kernel, we transform all training geometries into a canonical permutation $\vec{x}_i \equiv \mat{P}_{i1}\vec{x}_i$, enabling the use of uniform symmetry transformations $\mat{P}_{j} \equiv \mat{P}_{1j}$. Simplifying Eq.~\ref{eq:data_augmentation_model} accordingly, gives rise to the symmetric kernel that we originally set off to construct
\begin{equation}
\begin{aligned}
\hat{f}(\vec{x}) &= \sum^M_i \alpha_{i} \sum^S_q  k(\vec{x},\mat{P}_{q} \vec{x}_i) \\
&= \sum^M_i \alpha_{i} k_\text{sym}(\vec{x},\vec{x}_i) \text{,}
\label{eq:sym_model}
\end{aligned}
\end{equation}
and yields a model with the exact same number of parameters as the original, non-symmetric one. This ansatz is known as \emph{invariant integration} and frequently applied to symmetrize ML potentials~\cite{haasdonk2007invariant, Bartok2013, glielmo2018efficient}. However, our solution, motivated by the concept of permutation-inversion groups~\cite{Longuet-Higgins1963}, is able to truncate the sum over potentially hundreds of thousands permutations in the full symmetric group of the molecule to a few physically reasonable ones. We remark that this step is essential in making invariant integration practical beyond systems with five or six identical atoms (with $5! = 120$ and $6! = 720$ permutations, respectively). For example, the molecules benzene, toluene and azobenzene each only have 12 physically relevant symmetries, whereas the associated symmetric groups have orders $6!6!$, $7!8!$ and $12!10!2!$, respectively. Our multi-partite matching algorithm is therefore able to shorten the sum over $S$ in Eq.~\ref{eq:sym_model} by up to 15 orders of magnitude, without significant loss of accuracy.

The data-driven symmetry adaptation approach outlined above can be applied universally, but in particular to the energy-conserving force field kernel that we have derived in the previous section. It improves the data-efficiency of the model in proportion to the number of symmetries that are present in a molecules (see Table~\ref{tab:sgdml_relative_performance}) to the point that costly high-level coupled-cluster CCSD(T) calculations become viable as reference data. We have shown that such a model effectively allows converged MD simulations with fully quantized electrons and nuclei for molecuels with up to a few dozen atoms~\cite{gdml2, Sauceda2019}.

\begin{table}
  \centering
  \caption{Relative increase in accuracy of the sGDML$@$DFT vs. the non-symmetric GDML model: the benefit of a symmetric model is directly linked to the number of permutational symmetries in the system. All symmetry counts in this table include the identity transformation.}
  \label{tab:sgdml_relative_performance}
  \begin{tabular}{lrrr}
  \toprule
    \multirow{2}{*}{\textbf{Molecule}} & \multirow{2}{*}{\textbf{\# Sym. in $k_\text{sym}$}} & \multicolumn{2}{c}{\textbf{$\Delta$ MAE} [$\%$]}\\
    \cline{3-4}
    & & Energy & Forces \\
    \hline
    Benzene & 12 & -1.6 & -62.3\\
    Uracil & 1 & 0.0 & 0.0\\
    Naphthalene & 4 & 0.0 & -52.2\\
    Aspirin & 6 & -29.6 & -31.3\\
    Salicylic acid & 1 & 0.0 & 0.0\\
    Malonaldehyde & 4 & -37.5 & -48.8\\
    Ethanol & 6 & -53.4 & -58.2\\
    Toluene & 12 & -16.7 & -67.4\\
    Paracetamol & 12 & -40.7 & -52.9\\
    Azobenzene & 8 & -74.3 & -47.4\\
    \bottomrule
  \end{tabular}
\end{table}

\section{Conclusion}

Typically, the parametrization of ML potentials relies on the availability of large reference datasets to obtain accurate results, which prevents the construction of ML models using costly high-level \emph{ab initio} methods due to the exploding computational cost. In this chapter, we have shown how to overcome this restrictive requirement by informing the model with fundamental physical invariances and conservation laws. Not only does this make the models more data-efficient, it also guarantees that the incorporated physics are represented without artifacts. In particular, we have successively developed a theoretical framework for construction of ML potentials that include the full set of temporal and spatial symmetries of molecules. Homogeneity of time implies energy conservation and global spatial symmetries include rotational and translational invariance of the energy.

Using a generalization of GPs to vector-valued Hilbert spaces, we have defined a predictor that explicitly maps to energy conserving solutions and thus allows the simultaneous prediction of accurate forces and energies at the same time. We have then extended this model to additionally incorporate all relevant rigid space group symmetries as well as dynamic non-rigid symmetries. Typically, the identification of symmetries requires chemical and physical intuition about the system at hand, which is impractical in a ML setting. Through a data-driven multi-partite matching approach, we have automated the discovery of permutation matrices of molecular graph pairs in different permutational configurations and thus between symmetric transformations undergone within the scope of a dataset. This allowed us to define a compact symmetric model that can be parametrized from very small training datasets of just a few hundreds of examples, enabling the direct construction of flexible molecular force fields from expensive high-level ab initio calculations.

While the resulting sGDML model constitutes a substantial step towards enabling highly accurate and thus truly predictive MD simulations, there is a number of challenges that remain to be solved in terms of its applicability and scaling to larger molecular systems. For example, a fragmentation of large atomistic systems would allow scaling up and transferable predictions across different molecules with similar atom types. The well-separated inter- and intramolecular correlation scales within molecular solids suggest that a hierarchical decomposition is possible with limited degradation of prediction accuracy. Furthermore, advanced sampling strategies could be employed to combine forces from different levels of theory to minimize the need for computationally-intensive ab initio calculations even further.

\section{Data and Software}

The ML potentials described in this chapter are implemented in the sGDML software package -- see \texttt{www.sgdml.org} for details. Reconstruct FFs from your own datasets today!

\greybox{\textbf{Note:} This chapter is adapted with permission from Chmiela, 2019 \cite{chmiela2019thesis}.}

\bibliographystyle{unsrt}
\bibliography{clean}% Produces the bibliography via BibTeX.

\end{document}